\documentclass[prl,aps,twocolumn,showpacs,amsmath,amssymb,superscriptaddress]{revtex4}
\usepackage{graphicx}
\usepackage{dcolumn}
\usepackage{bm}

\newcommand{\braopket}[3]{ \left\langle #1 \right| #2 \left| #3 \right\rangle }
\newcommand{\ketbra}[2]{ | #1 \left\rangle \right\langle #2 |}

\begin{document}

\title{Spin-orbit coupling and Berry phase with ultracold atoms in 2D optical lattices}

\author{Artem M. Dudarev}
 \email{dudarev@physics.utexas.edu}
\affiliation{Department of Physics, The University of Texas,
Austin, Texas 78712-1081} \affiliation{Center for Nonlinear
Dynamics, The University of Texas, Austin, Texas 78712-1081}
\author{Roberto B. Diener}
\affiliation{Department of Physics, The Ohio State University, Columbus, 
Ohio 43210}
\affiliation{Department of Physics, The University of Texas,
Austin, Texas 78712-1081}
\author{Iacopo Carusotto}
\affiliation{Laboratoire Kastler Brossel, \'Ecole Normale Sup\'erieure, 24 rue Lhomond, 75231 Paris Cedex 05, France}
\affiliation{BEC-INFM Research and Development Center, I-38050 Povo, Trento, Italy}
\author{Qian Niu}
\affiliation{Department of Physics, The University of Texas,
Austin, Texas 78712-1081}

\date{\today}

\begin{abstract}
We show how spin-orbit coupling and Berry phase can appear in two-dimensional optical lattices by coupling atoms' internal degrees of freedom to radiation. The Rashba Hamiltonian, a standard description of spin-orbit coupling for two-dimensional electrons, is obtained for the atoms under certain circumstances. We discuss the possibility of observing associated phenomena, such as the anomalous Hall and spin Hall effects, with cold atoms in optical lattices.
\end{abstract}


\maketitle

In the last decade many experiments have studied the quantum motion of ultracold atoms in periodic potentials created by standing waves of light~\cite{raizen97}. Phenomena, such as Bloch oscillations and Wannier-Stark ladders, impossible to observe for electrons in metals due to short relaxation times, have been observed. 
Atoms in different internal (spin) states can interact with laser light in different ways depending on its polarization.  This effect has recently been applied to the experimental study of quantum transport of atoms in one-dimensional optical lattices in the {\it localized} regime~\cite{haycock00,mandel03}, for which a wide range of phenomena has been theoretically proposed~\cite{spin-theory}.
Here we consider ultracold atoms dynamics in two-dimensional (2D) optical lattices in the {\it itinerant} regime where motion through the lattice is of concern.

An intrinsically new aspect of a higher dimensional lattice is the possible appearance of a geometric phase in crystal momentum space~\cite{chang96,sundaram99}.  This can arise whenever time-reversal and/or spatial inversion symmetries are broken in the lattice, or in the presence of spin-orbit coupling.  In solid state systems, such as ferromagnetic crystals for example, the geometric phase is responsible for the anomalous Hall effect (AHE)~\cite{culcer03}, that is the generation of a transverse current by an electric field even in the absence of a magnetic field.  A closely related phenomenon, recently proposed for spintronics applications~\cite{wolf01}, is the spin Hall effect (SHE)~\cite{murakami-sinova-culcer03}, that is the production of a transverse spin current by an electric field in the presence of a significant spin-orbit coupling.  The observation of the spin Hall effect with present technology is however still an open challenge in solid state systems.   

In this Letter, we show how spin-orbit coupling and geometric phase may be produced for atoms in 2D optical lattices, and we propose experiments to explore their consequences.  Optical lattices provide great flexibility with which potentials can be created and atomic quantum states prepared. By choosing the polarization of the beams appropriately, the internal degrees of freedom of the atom can be coupled to their momenta as in the (relativistic) spin-orbit effect for electrons in solids. In particular, we show how a Hamiltonian similar to the Rashba Hamiltonian for electrons in 2D semiconductor systems~\cite{bychkov84} can arise in the description of atoms propagating in an optical lattice produced by the interference of suitably polarized laser beams. A constant force field, analogous to an electric field for electrons can be produced by accelerating the lattice~\cite{raizen97}.  By studying the transport of the atoms in these spin-dependent lattices, one can observe effects similar to AHE and SHE.

Let us consider the Hamiltonian for an atom interacting with a configuration of laser beams producing an electric field ${\bf E}({\bf r})$. In general, if the detuning of the light frequency $\omega$ from the resonance frequency is large with respect to the radiative width of the excited states, spontaneous emission is suppressed and we can adiabatically eliminate the excited states by writing an effective Hamiltonian which involves only the ground state sublevels~\cite{carusotto03,bfict}: 
\begin{eqnarray}
 H_{{\rm eff}} & = & 
 \sum\limits_{\alpha \beta \gamma } {\frac{\left( {{\bf E}^* ({\bf r}) \cdot {\bf D}_{\gamma \alpha }^* \left| {g:\alpha } \right\rangle  } \right)\left( {{\bf E}({\bf r}) \cdot {\bf D}_{\gamma \beta } \left\langle {g:\beta } \right|} \right)}{\hbar (\omega - \omega _\gamma)}}
 \nonumber \\ 
  & = & \sum_{\alpha \beta} V_{\alpha \beta}({\mathbf r})
\ketbra{g:\alpha}{g:\beta}.
\end{eqnarray}
For a ground state of total angular momentum $F$,  
the indexes $\alpha$ and $\beta$ run over the $2F+1$ Zeeman sublevels.
${\bf D}_{\gamma \beta}=\braopket{e:\gamma}{ {\bf d}}{g:\beta}$
is the dipole matrix element between the ground state sublevel $\beta$ and the excited state sublevel $\gamma$ (of energy $\hbar \omega _\gamma$).

Given the analogy with electrons, in the following we 
shall focus on the case of atoms with $F=1/2$. This case can be effectively realized using alkali atoms, cooling and manipulation of which have gone a long way. In particular,
$^6$Li atoms have a ground state hyperfine component of $F=1/2$.
Similar features are expected to occur for higher values of
the atomic spin $F$ as well, but we leave their detailed analysis for
the future. 

In the $F=1/2$ case, the external potential $V_{\alpha\beta}({\mathbf r})$ can be simply 
written in terms of a fictitious magnetic field ${\mathbf B}({\mathbf r})$ coupled
to the total atomic angular momentum operator ${\hat {\mathbf F}}$:
\begin{equation}
	V_{\alpha\beta}({\mathbf r})=V({\mathbf r}) \delta _{\alpha \beta}+{\mathbf B}({\mathbf
 r})\cdot
 {\hat {\mathbf F}}_{\alpha\beta}.
\label{eq:pot}
\end{equation}
The scalar potential $V({\mathbf r})$ is proportional to the local light 
intensity, while the vectorial field 
${\mathbf B}({\mathbf r})$ is proportional to the local electromagnetic spin: 
\begin{eqnarray}
V({\mathbf r})&=&b_0 {\mathbf E}^*({\mathbf r})\cdot {\mathbf
E}({\mathbf r}) 
\label{eq:V} \\
{\mathbf B}({\mathbf r})&=&-i b_1 {\mathbf E}^*({\mathbf r}) \times {\mathbf
E}({\mathbf r}). \label{eq:B}
\end{eqnarray}
The proportionality coefficients $b_{0,1}$ depend on the details of the atomic structure as well as on the light frequency.  As discussed in~\cite{carusotto03} an effective coupling to the fictitious magnetic field requires the detuning from the excited state to be smaller than the fine structure of the excited state.  Some consequences of the fictitious magnetic field have already been investigated from many different points of view, for instance Sysiphus cooling in resonant optical lattices~\cite{dalibard89}, or NMR type experiments~\cite{bfict-nmr}.  The fictitious magnetic field was crucial for the observation of mesoscopic quantum tunneling in one-dimensional spin-dependent optical lattices~\cite{haycock00}.

\begin{figure}[t]
\includegraphics[width=3.5in]{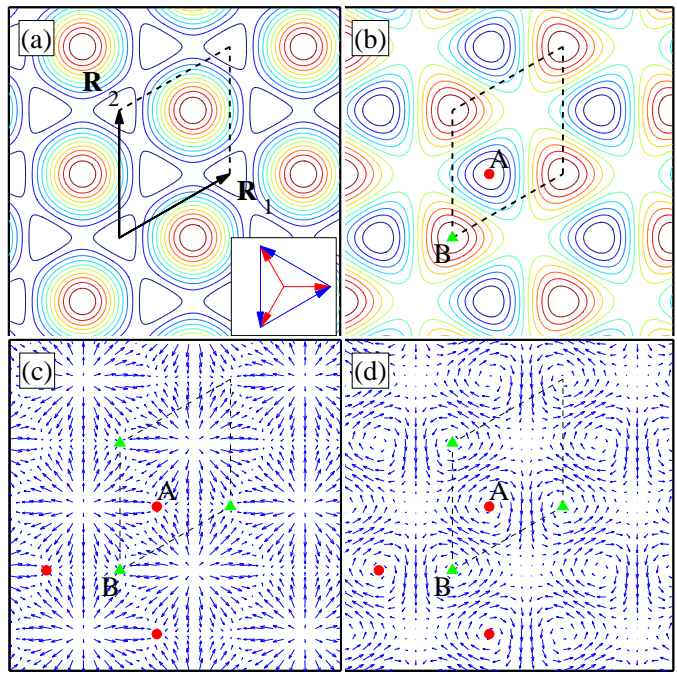}
\caption{\label{fig:pot} Spatial profile of the different terms of the potential~(\ref{eq:pot_sp}): (a) $V_0$ term, (b) $V_1$ term, (c) $V_2$ term, (d) $V_3 $ term. The amplitude of a vector component is proportional to the length of the arrow. The inset in panel (a) shows the configuration of the laser wave vectors ${\mathbf q}_i$ (red arrows) and the lattice vectors ${\mathbf k}_i$ (blue arrows). Red dots indicate wells A, green triangles indicate wells B. Dashed lines show a unit cell.}
\end{figure}

We consider a 2D optical lattice created with three laser beams propagating in a plane with equal angles between them (see Fig.~\ref{fig:pot}), a geometry already studied in the dissipative regime~\cite{grynberg93}. As done in~\cite{deutsch98}, we however restrict our attention on the dissipationless case. We choose the polarization of the electric fields in such a way that the components respectively parallel and perpendicular to the plane are the same for all three beams

\begin{equation}
		{\bf E}_i  = (\alpha \hat z + \beta \hat z \times {\bf \hat q}_i) e^{i{\bf q}_i {\bf r}};
\end{equation}
the ${\mathbf q}_i$ are the wave vectors of the light beams, coefficients
 $\alpha$ and $\beta$ are complex and ${\hat z}$ is the unit
 vector perpendicular to the plane. For this configuration, the scalar potential~(\ref{eq:V}) and the fictitious magnetic field~(\ref{eq:B}) are 
\begin{eqnarray}
	\label{eq:pot_sp}
		V({\bf r}) & = & \sum\limits_i {V_0 \cos {\bf k}_i {\bf r}},  \\ 
		{\bf B}({\bf r}) & = & \sum\limits_i {\left( {V_1 \hat z\sin {\bf k}_i {\bf r} + V_2 {\bf k}_i \sin {\bf k}_i {\bf r} + V_3 \hat z \times {\bf k}_i \cos {\bf k}_i {\bf r}} \right)}, \nonumber
\end{eqnarray}
where ${\bf k}_1 = {\bf q}_2 - {\bf q}_3$, {\it etc.}, and the
amplitudes $V_i$ are 
\begin{equation}
	\begin{array}{l}
		V_0  = b_0 \left( \left| \alpha  \right|^2  + \left| \beta  \right|^2 \cos (2\pi /3) \right), \\ 
		V_1  = 2 b_1 \left| \beta  \right|^2 \sin (2\pi /3), \\ 
		V_2  =  - 4 b_1 \sin (\pi /3){\mathop{\rm Re}\nolimits} \left( {\alpha ^* \beta } \right), \\ 
		V_3  = 2 b_1 {\mathop{\rm Im}\nolimits} \left( {\alpha ^* \beta } \right).
	\end{array}
\end{equation}
We will use $M$, $(q \sqrt{3})^{-1}$ and $6 E_r$ as the basic units of mass, length and energy, where $M$ is the mass of the atom, $q_i$ is the light wave vector, and $E_r = \hbar^2 q^2/2 M$ is the recoil energy.
Experimentally, potentials with depths up to $20 E_r$ are achievable in the dissipationless regime.

\begin{figure}[t]
	\begin{center}
		\includegraphics[width=3.5in]{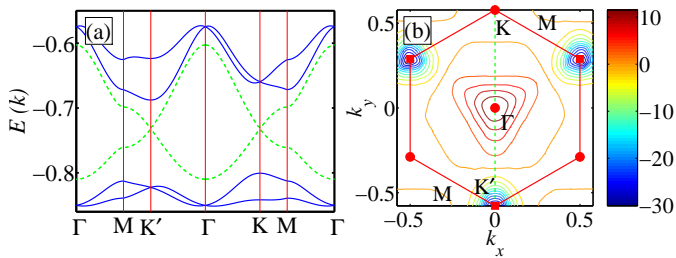}
	\end{center}
	\caption{(a) Dispersion of the lowest four bands in the potential~(\ref{eq:pot_sp}). The green dashed line corresponds to the case in which only a scalar potential is present, $V_0=1$. The blue solid line is for the case $V_0=1$, $V_1=V_3=0.1$ and $V_2=0$~\cite{exp-param}.
	(b) Berry curvature for the lowest band in the presence of an external field $h_0=0.005$. The band structure in the presence of $h_0$ is essentially the same as shown in (a), except for a splitting by $\approx 0.01$ at the degeneracy points $\Gamma$, K and ${\rm K^\prime}$.}
	\label{fig:disp_omega}
\end{figure}

In what follows we are interested in the dynamics of an atomic distribution prepared in the lowest band of the potential. The different terms in the Hamiltonian proportional to the $V_i$ are shown in Fig.~\ref{fig:pot}. We consider the case in which $\left| V_0 \right| \gg \left| V_1\right|,\left|V_2\right|,\left|V_3\right|$ ($\left| \alpha \right| \gg \left| \beta \right|$). Together with a direct numerical calculation of the band structure and Berry curvature (see Fig.~\ref{fig:disp_omega}), we consider a tight-binding model that allows us to obtain an effective Hamiltonian close to the $\Gamma$ point.  The scalar part of the potential creates a honeycomb lattice which is perturbed by the vector part. Introduction of the term with $V_1$ makes the lattice asymmetric for each spin component, similar to the one recently studied by us~\cite{diener03}. $V_2$ and $V_3$ lead to hopping processes with simultaneous spin flipping. The tight-binding approach results in the effective Hamiltonian
\begin{equation}
\left( {\begin{array}{*{20}c}
   {\epsilon _A^ \uparrow + h_0 } & v & 0 & {v_1 }  \\
   {v^* } & {\epsilon _B^ \uparrow + h_0 } & {v_2 } & 0  \\
   0 & {v_2^* } & {\epsilon _A^ \downarrow - h_0 } & v  \\
   {v_1^* } & 0 & {v^* } & {\epsilon _B^ \downarrow - h_0 }  \\
\end{array}} \right)\begin{array}{*{20}c}
   {\left| {A, \uparrow } \right\rangle }  \\
   {\left| {B, \uparrow } \right\rangle }  \\
   {\left| {A, \downarrow } \right\rangle }  \\
   {\left| {B, \downarrow } \right\rangle }  \\
\end{array}.
\label{eq:ham-matrix}
\end{equation}
Since lattices for
 different spins are mirror images of each other, the on-site energies
 for different spins are such that $\epsilon _A^ \uparrow = -\epsilon _B^ \uparrow = -\epsilon _A^ \downarrow = \epsilon _B^ \downarrow$. Off-diagonal matrix elements are 
\begin{equation}
	\begin{array}{l}
		v({\bf k}) =  - t\left( {1 + e^{i{\bf kR}_1 }  + e^{i{\bf kR}_2 } } \right), \\ 
		v_1 ({\bf k}) = \tilde t \left( {e^{ - i\pi /3}  + e^{i\pi /3} e^{i{\bf kR}_1 }  - e^{i{\bf kR}_2 } } \right), \\ 
		v_2 ({\bf k}) = \tilde t \left( {e^{ - i\pi /3}  + e^{i\pi /3} e^{ - i{\bf kR}_1 }  - e^{ - i{\bf kR}_2 } } \right),
	\end{array}
\end{equation}
where $t$ is proportional to the overlap of the wave functions in different sites with the same spin, $\tilde t$ is proportional to the overlap of wave functions at different sites with different spin, and ${\bf R}_i$ are the lattice primitive vectors (see Fig.~\ref{fig:pot}). The coefficient $t$ is determined by the scalar potential, while $\tilde t$ is governed mostly by $V_3$ terms (to the lowest order, terms with $V_2$ do not affect the dynamics, since the corresponding ${\bf B}$ field vanishes at the place where the overlap between on-site wave functions is the largest, see Fig.~\ref{fig:pot}). The parameter $h_0$ describes an additional external field.

For vanishing fictitious magnetic field ${\mathbf B}$ and external
field $h_0$, the bands have a two-fold spin degeneracy at all points
of the Brillouin zone; at points ${\rm K}$ and ${\rm K^\prime}$, where bands cross, the
degeneracy is four-fold.
As the $V_1$ term is added, the bands keep the two-fold spin
degeneracy, but the four-fold one at ${\rm K}$ and ${\rm K^\prime}$ is lifted and a gap is
correspondingly opened. Spin degeneracy is then lifted by the inclusion of the $V_3$ term.  The band degeneracies at the high symmetry points can be understood from the tight-binding model simply by looking at the properties of the off-diagonal matrix elements in (\ref{eq:ham-matrix}).  For instance, at the $\Gamma$ point one has $v_{1,2}=0$ and only $v\neq 0$, therefore the upper and lower bands are two-fold degenerate there. An effective Hamiltonian close to the $\Gamma$ point can therefore be obtained simply by treating $v_1$ and $v_2$ as a perturbation.

\begin{figure}[b]
	\begin{center}
		\includegraphics[width=3in]{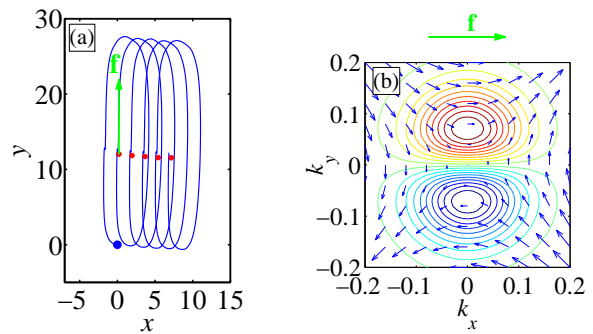}
	\end{center}
	\caption{
	(a) AHE. Trajectory of the wave packet center for a force ${\mathbf f}$ in the positive $y$ direction. Red dots are period-averaged positions. The initial position is indicated by the blue dot. 
	(b) SHE. Distribution of the spin vertical component as a function of momentum when finite force ${\mathbf f}=f_x {\hat x}$ is applied. Red is positive, blue is negative. The blue arrows show effective momentum dependent field $\Delta _{\bf k}$ close to the $\Gamma$ point.} 
	\label{fig:ahe-she}
\end{figure}

The spectrum of the unperturbed Hamiltonian in the absence of the external field consists of two doubly degenerate states of energies
\begin{equation}
\epsilon_\pm=\pm\frac{1}{2}\sqrt{\big(\epsilon_B^\uparrow-\epsilon_A^\uparrow\big)^2
+4|v({\mathbf k})|^2}.
\end{equation}
For small $k$ around the $\Gamma$ point, the dispersion is quadratic in $k$.  The corresponding two lowest eigenvectors are of the form:
\begin{equation}
	\begin{array}{c}
		\left| { \uparrow , - } \right\rangle  = a\left| {A, \uparrow } \right\rangle  + b\left| {B, \uparrow } \right\rangle,  \\ 
		\left| { \downarrow , - } \right\rangle  = b\left| {A, \downarrow } \right\rangle  + a\left| {B, \downarrow } \right\rangle.  \\ 
	\end{array}
\end{equation}
 In the subspace spanned by these two eigenvectors the effective Hamiltonian once $v_1$ and $v_2$ are taken into account has the form:
\begin{equation}
	\label{eq:ham-gamma}
	H_\Gamma  ({\bf k}) = \epsilon_ - I +h_0 \sigma _z - \gamma (k_y \sigma _x + k_x \sigma _y),
\end{equation}
where $\sigma _i$ are Pauli matrices and $I$ is the unit matrix.  
This reduces to the standard form of the Rashba Hamiltonian if one makes a 
global spin-rotation about the $\sigma_y$ axis to flip the signs of $\sigma_x$ and $\sigma_z$.
For the model here discussed, the value of the spin-orbit coupling parameter $\gamma=0.08$ can be extracted from the relative slope of the two lowest eigenstates. A Hamiltonian of such a form has been recently predicted to give both AHE~\cite{culcer03} and SHE~\cite{murakami-sinova-culcer03} in solid state systems.

The simple analytic form of the Hamiltonian in~(\ref{eq:ham-gamma}) gives a simple expression for the Berry curvature which is responsible for AHE~\cite{culcer03,chang96,sundaram99}
\begin{equation}
	\Omega _z^{ \uparrow / \downarrow }  =  \mp \frac{1}{2}\frac{{ \gamma ^2 h_0 }}{{\left( {h_0^2  + \gamma ^2 k^2 } \right)^{3/2} }}.
\end{equation}
The results of the Berry curvature calculation for the {\it continuous} potential in Eq.~(\ref{eq:pot_sp}) are shown in Fig.~\ref{fig:disp_omega}. For $h_0=0$ the Berry curvature is zero everywhere and singular at the points where the bands touch. Any finite value of the external field $h_0$ completely removes the degeneracy of the bands and makes the Berry curvature to be peaked around the $\Gamma$ and ${\rm K} ^\prime$ points with a finite maximum and width. An effect similar to AHE can therefore be observed with cold atoms: a wave packet initially prepared in the {\it lowest band} with a small quasimomentum spread $\Delta k \ll 1$~\cite{friedman02} around the $\Gamma$ point and accelerated in the $\Gamma$K direction performs Bloch oscillations in the direction of the drive and at the same time it drifts along the perpendicular direction because of the geometrical phase accumulated~\footnote{For this direction, symmetry arguments guarantee   that no transverse drift coming from the band asymmetry can occur.}.  In Fig.~\ref{fig:ahe-she} (a), we show the trajectory of the wave packet center calculated with semiclassical equations including the Berry curvature effects~\cite{chang96,sundaram99}. The external force has been taken as $f=0.001$. In order to maximise the drift in a given time, the largest force that preserves adiabatic evolution in the first band has to be chosen in an actual experiment.
The magnitude of the acceleration creating the force ${\bf f}$ is $\sim 500 {\rm ~ m/s^2}$, a value already demonstrated in the framework of optical lattices~\cite{raizen97}. The chosen value of external field $h _0$ corresponds to $\sim 1 {\rm ~ mG}$. To observe a wave packet drift of $\sim 10 \lambda$, the acceleration needs to be applied for $\sim 50 {\rm ~ ms}$.

To observe an effect similar to SHE with cold atoms, a {\it spin-sensitive} measurement of the wave packet momentum distribution after it was exposed to an external force has to be performed. The quasimomentum spread has to be small enough, and the force applied for a short enough time, so that the wave packet stays within the region where~(\ref{eq:ham-gamma}) is valid. This region has a radius of the order of 0.2 around the $\Gamma$ point. As discussed in~\cite{murakami-sinova-culcer03} off-diagonal terms in Eq.~(\ref{eq:ham-gamma}) may be thought of as a momentum dependent magnetic field $\Delta _{\bf k} = - \gamma (k_y \hat x + k_x \hat y)$. When $h_0 = 0$ and an external force is applied, the spins, initially parallel to the plane, acquire a non-vanishing $z$-component, i.e. ``tilt vertically''. For an atom with quasimomentum $(k_x,k_y)$ and a force along the ${\hat x}$ direction ${\mathbf f}=f_x {\hat x}$, the $z$-component of the spin is:
\begin{equation}
	n_{z,{\bf k}} = \frac{f _x k _y}{\gamma k^3}.
\end{equation}
The atoms with a positive $k_y$ thus acquire a positive $z$-component of the spin, while the atoms with a  negative $k_y$ acquire a negative one. For the momentum spread, we have taken $\Delta k_y=0.1$, which
 corresponds to the minimum of the first band when $h_0=0$, and for
 the force $f=10^{-4}$ ($\approx 50\textrm{m}/\textrm{s}^2$). For these
 values, we predict $n_{z,{\mathbf k}}=0.125$. To measure the spin tilting in the vertical direction one should measure the momentum distribution of the different vertical spin components after the force has been applied. Spin dynamics can be frozen by suddenly switching on an external $h_0 \gg \gamma \Delta k$.  The lattice potential is then adiabatically removed: this transforms the quasimomentum distribution into a true momentum one so that SHE can be observed as the motion of the different spin $z$-components in opposite directions along the $y$ axis.
 
 In conclusion, we have considered continuing the parallels between the dynamics of electrons in solids and cold atoms in optical  potentials. We have shown that the coupling of the atomic internal degrees of freedom to the translational ones can be described in terms of the same Rashba Hamiltonian as used to model spin-orbit coupling in solids, and we have predicted the possibility of observing AHE and SHE with cold atoms.  With respect to solid state systems, atomic ones are expected to provide a much cleaner environment and a easier tunability of the system parameters. This will allow for a deeper understanding of the basic physical mechanisms that may be important for future spintronics applications.

We acknowledge support from the NSF, the R.A. Welch Foundation,
and discussions with M.G. Raizen.
IC and QN acknowledge hospitality at the Aspen Center for Physics.


\begin{thebibliography}{99}
\bibitem{exp-param} The values of $V_i$ for $^6$Li correspond to almost circular polarization $\left| \alpha \right| / \left| \beta \right| = 0.87$, $\phi _\alpha - \phi _\beta = \pi / 2$; detuning from the resonance $\Delta \sim 450$ GHz; intensity of the beams $I \sim 70$ mW$/$cm$^2$. In this case scattering rate for a free atom is $\Gamma _{sc} \sim 200$ s$^{-1}$ and even much smaller for the ones trapped in the lowest band.
\end{thebibliography}
\end{document}